\documentclass{nature_modified}
\usepackage{hyperref}
\usepackage{graphicx}
\usepackage{multibib}
\newcites{Methods}{References}

\title{A seven-planet resonant chain in TRAPPIST-1}
\author{Rodrigo Luger$^{1,2}$, 
Marko Sestovic$^{3}$, 
Ethan Kruse$^{1}$, 
Simon L. Grimm$^{3}$, 
Brice-Olivier Demory$^{3}$, 
Eric Agol$^{1,2}$,
Emeline Bolmont$^{4}$, 
Daniel Fabrycky$^{5}$,
Catarina S. Fernandes$^{6}$,
Val{\'e}rie Van Grootel$^{6}$, 
Adam Burgasser$^{7}$, 
Micha\"el Gillon$^{6}$, 
James G. Ingalls$^{8}$, 
Emmanu\"el Jehin$^{6}$,
Sean N. Raymond$^{9}$, 
Franck Selsis$^{9}$,
Amaury H.M.J. Triaud$^{10}$, 
Thomas Barclay$^{11,12,13}$, 
Geert Barentsen$^{11,12}$,
Steve B. Howell$^{11}$,
Laetitia Delrez$^{6,14}$, 
Julien de Wit$^{15}$, 
Daniel Foreman-Mackey$^{1,23}$,
Daniel L. Holdsworth$^{16}$, 
J{\'e}r{\'e}my Leconte$^{9}$, 
Susan Lederer$^{17}$,  
Martin Turbet$^{18}$,
Yaseen Almleaky$^{19,20}$, 
Zouhair Benkhaldoun$^{21}$, 
Pierre Magain$^{6}$, 
Brett Morris$^{1}$,
Kevin Heng$^{3}$,
\& Didier Queloz$^{14,22}$
}

\begin{document}

\maketitle

\begin{affiliations}
%
% 1. Luger, Agol, Kruse, Foreman-Mackey:
\item Astronomy Department, University of Washington, Seattle, WA, 98195, USA
%
% 2. Luger, Agol:
\item NASA Astrobiology Institute's Virtual Planetary Laboratory, Seattle, WA, 98195, USA
%
% 3. Sestovic, Grimm, Demory, Kevin Heng:
\item University of Bern, Center for Space and Habitability, Sidlerstrasse 5, CH-3012, Bern, Switzerland
%
% 4. Bolmont:
\item Laboratoire AIM Paris-Saclay, CEA/DRF - CNRS - Universit{\'e} Paris Diderot, IRFU/SAp Centre de Saclay, 91191, Gif-sur-Yvette
%
% 5. Fabrycky:
\item Department of Astronomy and Astrophysics, University of Chicago, 5640 South Ellis Avenue, Chicago, IL 60637, USA
%
% 6. Grootel, Gillon, Fernandes, Jehin:
\item Space Sciences, Technologies and Astrophysics Research (STAR) Institute, Universit\'e de Li\`ege, All\'ee du 6 Ao\^ut 19C, B-4000 Li\`ege, Belgium
%
% 7. Burgasser:
\item Center for Astrophysics and Space Science, University of California San Diego, La Jolla, CA, 92093, USA
%
% 8. Ingalls:
\item IPAC, Mail Code 314-6, California Institute of Technology, 1200 E California Boulevard,  Pasadena, CA 91125, USA
%
% 9. Selsis, Leconte, Raymond:
\item Laboratoire d'Astrophysique de Bordeaux, Univ. Bordeaux, CNRS, B18N, all{\'e}e Geoffroy Saint-Hilaire, 33615 Pessac, France
%
% 10. Triaud:
\item Institute of Astronomy, Madingley Road, Cambridge CB3 0HA, UK
%
% 11. Barclay, Barentsen
\item NASA Ames Research Center, Moffett Field, CA 94043, USA
%
% 12. Barclay, Barentsen
\item Bay Area Environmental Research Institute, 25 2nd St. Ste 209 Petaluma, CA 94952, USA
%
% 13. Barclay
\item NASA Goddard Space Flight Center, Greenbelt, MD 20771, USA
%
% 14. Delrez, Queloz:
\item Cavendish Laboratory, J J Thomson Avenue, Cambridge, CB3 0HE, UK
%
% 15. de Wit:
\item Department of Earth, Atmospheric and Planetary Sciences, Massachusetts Institute of Technology, 77 Massachusetts Avenue, Cambridge, MA 02139, USA
%
% 16: Holdsworth:
\item Jeremiah Horrocks Institute, University of Central Lancashire, Preston PR1 2HE, UK
%
% 17: Lederer
\item NASA Johnson Space Center, 2101 NASA Parkway, Houston, Texas, 77058, USA
%
% 18. Turbet:
\item Laboratoire de M\'et\'eorologie Dynamique, Sorbonne Universit\'es, UPMC Univ Paris 06, CNRS, 4 place Jussieu, 75005 Paris, France
%
% 19. Almleaky:
\item Space and Astronomy Department, Faculty of Science, King Abdulaziz University, 21589 Jeddah, Saudi Arabia
%
% 20. Almleaky:
\item King Abdullah Centre for Crescent Observations and Astronomy (KACCOA), Makkah Clock, Saudia Arabia
%
% 21. Zouhair:
\item LPHEA laboratory, Ouka\"imeden Observatory, Cadi Ayyad University FSSM, BP 2390, Marrakesh, Morocco
%
% 22. Queloz:
\item Observatoire de Gen\`eve, Universit\'e de Gen\`eve, 51 chemin des Maillettes, CH-1290 Sauverny, Switzerland
%
% AND LAST BUT NOT LEAST
% 23. Foreman-Mackey:
\item Sagan Fellow
\end{affiliations}

%- Bold paragraph (150 words, need to be edited for clarity)

\begin{abstract}
The TRAPPIST-1 system is the first transiting planet system found orbiting an ultra-cool dwarf star\cite{Gillon2016}. At least seven planets similar to Earth in radius were previously found to transit this host star\cite{Gillon2017}.  Subsequently, TRAPPIST-1 was  observed as part of the {\it K2} mission and, with these new data, we report the measurement of an \mbox{18.77\,d} orbital period for the outermost transiting planet, \mbox{TRAPPIST-1h}, which was unconstrained until now. This value matches our theoretical expectations based on Laplace relations\cite{Quillen:2011} and places \mbox{TRAPPIST-1h} as the seventh member of a complex chain, with three-body resonances linking every member. We find that \mbox{TRAPPIST-1h} has a radius of 0.727\,R$_{\oplus}$ and an equilibrium temperature of 169\,K. We have also measured the rotational period of the star at 3.3\,d and detected a number of flares consistent with a low-activity, middle-aged, late M dwarf.
\end{abstract}

%- Main text (~2000 words)

%-  Description of the K2 observations (Brice)
The star TRAPPIST-1 (EPIC 246199087) was observed for 79 days by NASA's {\it Kepler} Space Telescope in its two-reaction wheel mission\cite{Howell:2014} ({\it K2}) as part of Campaign 12, starting on 2016 Dec 15 and ending on 2017 Mar 04. The spacecraft was in safe mode between 2017 Feb 1 and 2017 Feb 6, resulting in a 5-day data loss. Typically upon downlink from the spacecraft, the raw cadence data are calibrated with the {\it Kepler} pipeline\cite{Quintana:2010}, a lengthy procedure that includes background subtraction, smear removal, and undershoot and nonlinearity corrections. However, given the unique science drivers in this dataset, the raw, uncalibrated data for Campaign 12 were made publicly available on 2017 Mar 8 shortly after downlink. We download and calibrate the long cadence ($t_\mathrm{exp} = 30\,\mathrm{min}$) and short cadence ($t_\mathrm{exp} = 1\,\mathrm{min}$) light curves using a simple column-by-column background subtraction, which also removes smear and dark noise (see Methods). 
%-  Description of the data reduction/analysis. (Rodrigo, Marko, Ethan)
Because of its two failed reaction wheels, rolling motion of the {\it Kepler} spacecraft due to torque imbalances introduces strong instrumental signals, leading to an increase in photometric noise by a factor of ${\sim}3-5$ compared to the original mission. Since TRAPPIST-1 is a faint M8 dwarf with {\it Kepler} magnitude $Kp \sim 16-17$ (see Methods), these instrumental signals must be carefully removed to reach the $\sim$0.1\% relative photometric precision required to detect Earth-size transits\cite{Demory:2016}. To this end, we detrend the long cadence light curve for TRAPPIST-1 using both \texttt{EVEREST}\cite{Luger2016,Luger2017} and a Gaussian process-based pipeline, achieving an average 6-hr photometric precision of 281.3\,ppm, a factor of 3 improvement over the raw light curve. After analysis of the long cadence light curve, we detrend the short cadence light curve in the vicinity of features of interest, achieving comparable or higher 6-hr precision (see Methods).

%-  Results (900 words) =============

%   - 1. planet search results
% A. Rodrigo's first analysis
We conduct three separate transit searches on the long cadence light curve, aiming to constrain the period of TRAPPIST-1h, which had only been observed to transit once\cite{Gillon2017}, as well as to detect additional planets in the system. 
A dynamical analysis made by our team prior to the release of the {\it K2} data suggested certain values of the period of TRAPPIST-1h based on the presence of three-body resonances among the planets.  
Three-body resonances satisfy $pP_1^{-1}-(p+q)P_2^{-1}+qP_3^{-1} \approx 0$ and $p\lambda_1-(p+q)\lambda_2+q\lambda_3 = \phi$ for integers $p, q$ and where $P_i$ and $\lambda_i$ are the period and mean longitude of the $i^{\rm th}$ planet\cite{Peale2002, Rivera2010} and $\phi$ is the 3-body angle which librates about a fixed value.
Such resonances occur both in our Solar System --- the archetypical case being the Laplace resonance among Jupiter's satellites, satisfying $(p,q)=(1,2)$ --- and in exoplanet systems, two of which were recently observed to have resonant chains among four planets: Kepler-223\cite{mills16} with $(p,q)=(1,1)$ and Kepler-80\cite{macdonald16} with $(p,q)=(2,3)$.
%
% DF reversed the roles of p and q from the emails. Now p multiplies the innermost planet's frequency, q multiplies the outermost planet's frequency, and -(p+q) multiplies the middle planet's frequency. This follows the convention set by MacDonald et al. 2016 (though beware Kepler-80's planets have an order going out, f,d,e,b,c!).
%
Among the inner six planets in TRAPPIST-1, there are four adjacent sets of three planets that satisfy this relation for $1\le p\le2$ and $1\le q\le3$ (Table~\ref{tab:3body}). This suggested that the period of planet TRAPPIST-1h may also satisfy a three-body resonance with TRAPPIST-1f and g. The six potential periods of TRAPPIST-1h that satisfy three-body relations with $1\le p,q\le 3$ are 18.766 d ($p=q=1$), 14.899 d ($p=1,q=2$), 39.026 d ($p=2,q=1$), 15.998 d ($p=2,q=3$), 13.941 d ($p=1,q=3$), and 25.345 d ($p=3,q=2$).  We examined $\sim 1000$ hours of ground-based data taken prior to the Spitzer dataset\cite{Gillon2017} and found a lack of obvious additional transits at the expected times for all of these periods save 18.766 d.
%(although the ground-based data show more variability, so there exists the possibility that some prior transits were missed due to noise). 
The period of 18.766 d corresponds to prior transit times in windows that were missed by the previous ground-based campaigns; hence, this was the only period that could not be ruled out. Furthermore, as this value is consistent with the period estimate of $20^{+15}_{-6}$\,d based on the duration of the {\it Spitzer} transit, we had reason to believe it was the correct period for TRAPPIST-1h.
To test this hypothesis, in our first transit search we simply fold the long cadence light curve at the four expected times of transit given this period and the single {\it Spitzer} transit time, finding evidence for a transiting planet at that period. Follow-up with detrended short cadence data confirms the transit-like shape of each of the four events and a depth consistent with that of TRAPPIST-1h (see Methods).
%

% B. Brice's analysis
To prove the uniqueness of this detection, in a second analysis, we search the detrended K2 light curve after subtracting a transit model including all known transits of planets b--g, based on published ephemerides and planet parameters\cite{Gillon2017}. We use the photometric residuals as input to a Box-fitting Least Squares (BLS) algorithm (see Methods) to search for additional transit signals. In this search, we do not impose prior information on TRAPPIST-1h. We find a periodic signal at $\sim18.77$\,d with a transit center at $\mathrm{BJD} = 2,457,756.39$, which matches the single transit observed by {\it Spitzer}\cite{Gillon2017}.

% C. Rodrigo's joint instrumental/transit search
Independently, we perform a joint instrumental/transit model fit to the data after subtracting a model for planets TRAPPIST-1b to g based on the {\it Spitzer} parameters (see Methods). We compute the relative likelihood (delta-chi squared) of a transit model with the best-fit {\it Spitzer} parameters of TRAPPIST-1h centered at every long cadence and sum the delta-chi squared values at the transit times corresponding to different periods in the range $1-50$\,d. Strong peaks emerge at $18.766$\,d and its aliases, corresponding to four transit-like events consistent with the parameters of TRAPPIST-1h at the times recovered in the previous searches.

%   - 2. Planetary signals and properties (600 words). (Laetitia, Brice, Michael)
We use the orbital period of TRAPPIST-1h determined in the previous step along with the parameters\cite{Gillon2017} of planets TRAPPIST-1b to g to determine whether a model including TRAPPIST-1h is favoured. This is achieved through Markov Chain Monte Carlo model fits with and without TRAPPIST-1h. We find a Bayes Factor of 90 in favour of a model that includes TRAPPIST-1h (see Methods), supporting the photometric detection of this seventh planet in the {\it K2} dataset.
%The orbital period of TRAPPIST-1h was determined to be 18.765\,d, consistent with our expectations that the planet would occupy a Laplace configuration. We are confident in our identification of planet $h$ owing to the detection of two clear photometric signals whose depth and a width are consistent with {\it Spitzer} observations (see Fig.~\ref{}), occurring at times that had been predicted and are consistent with the {\it Spitzer} ephemeris \footnote{A {\it Spitzer} observational window was scheduled for 2017-03-28 [check], during which planet $h$ will transit. This request was made prior to the {\it K2} data being sent to Earth for analysis.}.
% probably have to remove the following, given the recent TRAPPIST measurements that match Filippazzo 2015 (cf. emails from Valerie and Emmanuel)
%We use a new parallax measurement\cite{Weinberger2016}, yielding a larger distance of $12.64\pm0.18$ pc, which results in stellar radius/luminosity/mass larger by 4.5/9.1/14\% (respectively) relative to the values reported in the discovery paper\cite{Gillon2017}.  Consequently, the planet masses, incident fluxes, and radii are also larger by 4.5/9.1/14\%, while their inferred densities remain unaffected.
The detection of TRAPPIST-1h is thus supported by 1) the three transit search analyses that recovered both the orbital phase from the Spitzer ephemeris\cite{Gillon2017} and the period of 18.766\,d, 2) the Bayes Factor in favour of the 7-planet model and 3) the orbital period that is the exact value predicted by Laplace relations.  Figure~\ref{fig:fig1-hires} shows the full light curve, the newfound transits of TRAPPIST-1h, as well as an update to the geometry of the orbits given the new orbital period. In Table~\ref{tab:trappist1params} we report the properties of the planet derived in this study.

%-  Discussion (700 words) ==============

%   - Dynamics of the system (resonances, etc.) (300 words). (Eric, Sean, Brice)

%We use this orbital period determination for TRAPPIST-1h to analyze the dynamical configuration of the seven planet system, focusing on its intricate chain of mean motion resonances. We find that all sets of three adjacent planets in the TRAPPIST-1 system satisfy three-body resonances, as given in Table~\ref{tab:3body}.
%% RL removed this: we now introduce resonances above
%Two planets are in mean motion resonance if the ratio of their orbital periods is a ratio of small integers. This is nearly the case for each pair of adjacent planets in the TRAPPIST-1 system; for example, planets $d$ and $e$ have a period ratio of ${\sim}$2:3. 
%True resonances can be demonstrated by showing the libration of resonant angles related to orbital alignment\cite{murray99}, but with only transit data, the orientation of the orbits' semimajor axes is uncertain and therefore observing such a two-planet resonance is infeasible.

To characterize the three-body resonance, we use the transit timing data to identify $\phi$ (see Methods) for each set of three planets. A full transit-timing cycle has not elapsed within the data, so we cannot estimate the libration center for each $\phi$. However, we report in Table~\ref{tab:3body} the values of $\phi$ represented in the dataset. In the case of Jupiter's satellites, $\phi=180^\circ$, but due to the complexity of the multi-planetary system, we make no predictions for TRAPPIST-1 at this time. Migration and damping simulations applied to Kepler-80\cite{macdonald16} naturally predicted the values of the libration centers in that system; the measured values for TRAPPIST-1b--h call for future theoretical work interpreting them.

%   - Constraints on formation (200 words). (Sean)

The resonant structure of the system suggests that orbital migration may have played a role in its formation. Embedded in gaseous planet-forming disks, planets growing above $\sim 1$\,M$_\mathrm{Mars}$ create density perturbations that torque the planets' orbits and trigger radial migration\cite{baruteau14}. One model for the origin of low-mass planets found very close to their stars proposes that Mars- to Earth-sized planetary embryos form far from their stars and migrate inward\cite{raymond08}. The inner edge of the disk provides a migration barrier\cite{masset06} such that planets pile up into chains of mean motion resonances\cite{terquem07,ogihara09,cossou14}. This model matches the observed period ratio distribution of adjacent super-Earths\cite{fabrycky14} if the vast majority ($\sim 90\%$) of resonant chains become unstable and undergo a phase of giant impacts\cite{izidoro17}. Some resonant chains do survive, and a handful of multiple-resonant super-Earth systems have indeed been characterized\cite{godz16,mills16}. The TRAPPIST-1 system may thus represent a pristine surviving chain of mean motion resonances. Given that TRAPPIST-1's planet-forming disk was likely low in mass\cite{pascucci16} and the planets themselves are low-mass, their migration was likely relatively slow. This may explain why TRAPPIST-1's resonant chain is modestly less compact than chains in systems with more massive planets\cite{godz16,mills16,izidoro17}, which may have protected it from instability\cite{matsumoto12}. 

%   - Brief implications for tides/climates, etc. similar to what was done for the other T1 planets (200 words). (Jeremy, Franck, Emeline)

Given how close the planets orbit TRAPPIST-1, tidal interactions are likely to be important in the planets' orbital evolution\cite{ferrazmello08}. Tidal simulations of the system (see Methods) show that within a few Myr the eccentricity of each planet is damped to less than 0.01. Nonetheless, tidal heating is significant: all planets except TRAPPIST-1f and h have a tidal heat flux higher than Earth's total heat flux.  

The incident stellar flux on planet TRAPPIST-1h, 200\,W\,m$^{-2}$, is below the 300\,W\,m$^{-2}$ required to sustain surface liquid water under a N$_2$-CO$_2$-H$_2$O atmosphere (see Methods). 
%
% RL: Moved this to Methods
%The minimum stellar flux required for liquid water was calculated with the LMD 1D/3D Global Climate Model\cite{Turbet2016} using a TRAPPIST-1 spectrum and is in agreement with habitable zone boundaries computed for a 3000\,K star\cite{Kopparapu2013}. 
%
To obtain the missing 100\,W\,m$^{-2}$ from tidal heating would require a  high eccentricity strictly incompatible with the orbits of the other planets.
%[*** add something about a possible internal ocean in case of a water-rich planet ? *** ]
Our simulations show that the stellar input is also too low to sustain a thick
CO$_2$ atmosphere due to CO$_2$ condensation. In particular, CO$_2$ levels cannot exceed 100\,ppm within a 1\,bar N$_2$ atmosphere.
Alternatively, a liquid water ocean is possible under a layer of ice. The minimum thickness $h$ of this layer depends on the internal heat flux $\Phi_{int}$:
$$ h \sim 250 \left(\frac{\phi_{int}}{1\,\textrm{W\,m}^{-2}}\right)^{-1} ~\textrm{m} $$
Assuming the Earth's current geothermal flux, a layer of 2.8\,km (the mean depth of Earth's oceans) would be necessary.
%%%
%%%

% We need some kind of transition from talking about the planet to the star. 
While the long spin-down times of ultra-cool dwarfs prevent derivation of a robust gyrochronology relation\cite{reid2013new}, the rotational period of TRAPPIST-1 can be used to derive a provisional age estimate for the system.  Fourier analysis of the detrended {\it K2} data (Fig. \ref{fig:K2_detrended}), which is visibly modulated by star spots, leads to the determination of a rotational period of ${\sim} 3.3$\,d for the host star (see Methods). This rotation corresponds to an angular momentum of about 1\% that of the Sun. It is roughly in the middle of the period distribution of nearby late M dwarfs\cite{newton2016}, suggesting an age in the range $3-8$\,Gyr based on a star formation history that declines slightly with time\cite{aumer2009}. Such an age is consistent with the star's solar metallicity\cite{Gillon2016} and borderline thin disk/thick disk kinematics\cite{burgasser2015}. The amplitude of the modulation due to star spots and infrequent weak optical flares (0.26\,d$^{-1}$ for peak fluxes above 1\% of the continuum, 40 times less frequent than active M6-M9 dwarfs\cite{hilton2011}) are consistent with a low-activity M8 star, also arguing in favor of a relatively old system. Near the end of the {\it K2} campaign, a very energetic flare erupted, and was observed by {\it Kepler}. Full modeling of flares will be presented in a forthcoming paper. 

% EA - Here is my first stab at a summary paragraph:
% AJB - DO WE NEED IT?
% SML: I like this summary - gives good context with previous obs and what K2 did for us.

The {\it K2} observations of the TRAPPIST-1 star have enabled the detection of the orbital period of TRAPPIST-1h, continuing the pattern of Laplace resonance amongst adjacent triplets of planets.
%this pattern was in fact used to predict the period of TRAPPIST-1h prior to the {\it K2} campaign. 
We search for but do not detect additional planets in the system. Compared with previous ground-based and {\it Spitzer} observations, the continuous coverage, high precision, and shorter wavelength of the {\it K2} observations enable a robust estimate of the rotational period and flare activity of the star, motivating further study of the atmospheres and dynamical evolution of the planetary system.

%Gyrochronology is not calibrated for ultra-cool dwarfs, for which activity does not appear to be correlated with rotation\cite{reid2013new}. The moderate activity does not point towards a very young age. The evolution of magnetic activity of ultra-cool dwarfs is a largely unexplored field.

%TRAPPIST-1 appears to be a standard M8 star in terms of stellar variability.

%These two characteristics, unambiguously determined by continuous space-based observations, do not help to constrain the age of the system. Such a rotation period is relatively usual for ultra-cool dwarfs, and is rather slow by stellar rotation standards, with about 1\% of the angular momentum of the Sun. Gyrochronology is not calibrated for ultra-cool dwarfs, for which activity does not appear to be correlated with rotation\cite{reid2013new}. The moderate activity does not point towards a very young age. The evolution of magnetic activity of ultra-cool dwarfs is a largely unexplored field.

%%%
%%%
%%%
%%% REFERENCES FOR MAIN TEXT
%%%
%%%
%%%

\clearpage
\subsection{References}
\bibliographystyle{naturemag}
\bibliography{k2_trappist1}

%%%
%%%
%%%
%%% ADDENDUM: ACKNOWLEDGEMENTS, AUTHOR CONTRIBUTIONS, AUTHOR INFO
%%%
%%%
%%%

\begin{addendum}

\item[Acknowledgments] This paper includes data collected by the {\it K2} mission. Funding for the {\it K2} mission is provided by the NASA Science Mission directorate. This research has made use of NASA's Astrophysics Data System, the SIMBAD database and VizieR catalog access tool operated at CDS, Strasbourg, France. The data presented in this paper were obtained from the Mikulski Archive for Space Telescopes (MAST). R.L. and E.A. acknowledge support from NASA grant NNX14AK26G and from the NASA Astrobiology Institute's Virtual Planetary Laboratory Lead Team, funded through the NASA Astrobiology Institute under solicitation NNH12ZDA002C and Cooperative Agreement Number NNA13AA93A. E.A. acknowledges support from NASA grant NNX13AF62G and NSF grant AST-1615315. E.K. acknowledges support from an NSF Graduate Student Research Fellowship. B.-O.D. acknowledges support from the Swiss National Science Foundation in the form of a SNSF Professorship (PP00P2-163967). D.L.H. acknowledges financial support from the UK Science and Technology Facilities Council. M.S. and K.H. acknowledge support from the Swiss National Science Foundation.
A.J.B.\ acknowledges
funding support from the National Science Foundation under
award No.\ AST-1517177 and the National Aeronautics and Space Administration under Grant No.\ NNX15AI75G.
J. L. acknowledges funding from the European Research Council (ERC) under the European Union’s Horizon 2020 research and innovation programme (grant agreement No. 679030/WHIPLASH). M.G., E.J., and V.V.G. are F.R.S.-FNRS Research Associates.  S.N.R. thanks the Agence Nationale pour la Recherche for support via grant ANR-13-BS05-0003-002 (grant MOJO). The research leading to these results has received funding from the ERC under the FP/2007-2013 ERC Grant Agreement n° 336480 and from the ARC grant for Concerted Research Actions, financed by the Wallonia-
Brussels Federation.

\item[Author contributions] R.L. and M.S. led the detrending efforts with \texttt{EVEREST} and the GP-based pipeline, with inputs from E.A., J.G.I, E.K.\ and D.F.M. R.L.\ performed the preliminary manual search for transits of TRAPPIST-1h and the delta-chi squared search, with input from E.A, E.K.\ and D.F.M. S.L.G.\ took care of the K2 data handling. B.-O.D.\ led the collaboration, wrote the K2 proposal and performed an independent transit search and MCMC analysis of the K2 dataset. E.A.\ and D.F. led the dynamics and architecture of the system with inputs from S.N.R.\ and B.-O.D. E.B.\ took care of the tidal simulations. C.S.F, V.v.G., A.B., D.L.H.\ and B.M.\ conducted the work on stellar properties and variability and determined the stellar rotation period. S.N.R.\ led the formation and migration section. F.S., J.L.\ and M.T.\ worked on the atmospheric nature of TRAPPIST-1h. G.B.\ and T.B.\ helped with the handling of the uncalibrated K2 fits files. Figures were prepared by R.L., A.T., J.G.I., E.B\ and E.K. M.G., E.J., A.T., L.D., J.d.W, S.L., Y.A., Z.B., P.M., K.H.\ and D.Q.\ contributed to the discovery and characterisation of the TRAPPIST-1 system. All authors participated in the writing and commented on the paper.

\item[Correspondence] Correspondence and requests for materials should be addressed to R.L. (rodluger@uw.edu).

\item[Competing interests] The authors declare that they have no competing financial interests.

\end{addendum}

%%%
%%%
%%%
%%% MAIN FIGURES AND TABLES
%%%
%%%
%%%

\pagebreak

\begin{figure}
    \centerline{\includegraphics[width=0.6\hsize]{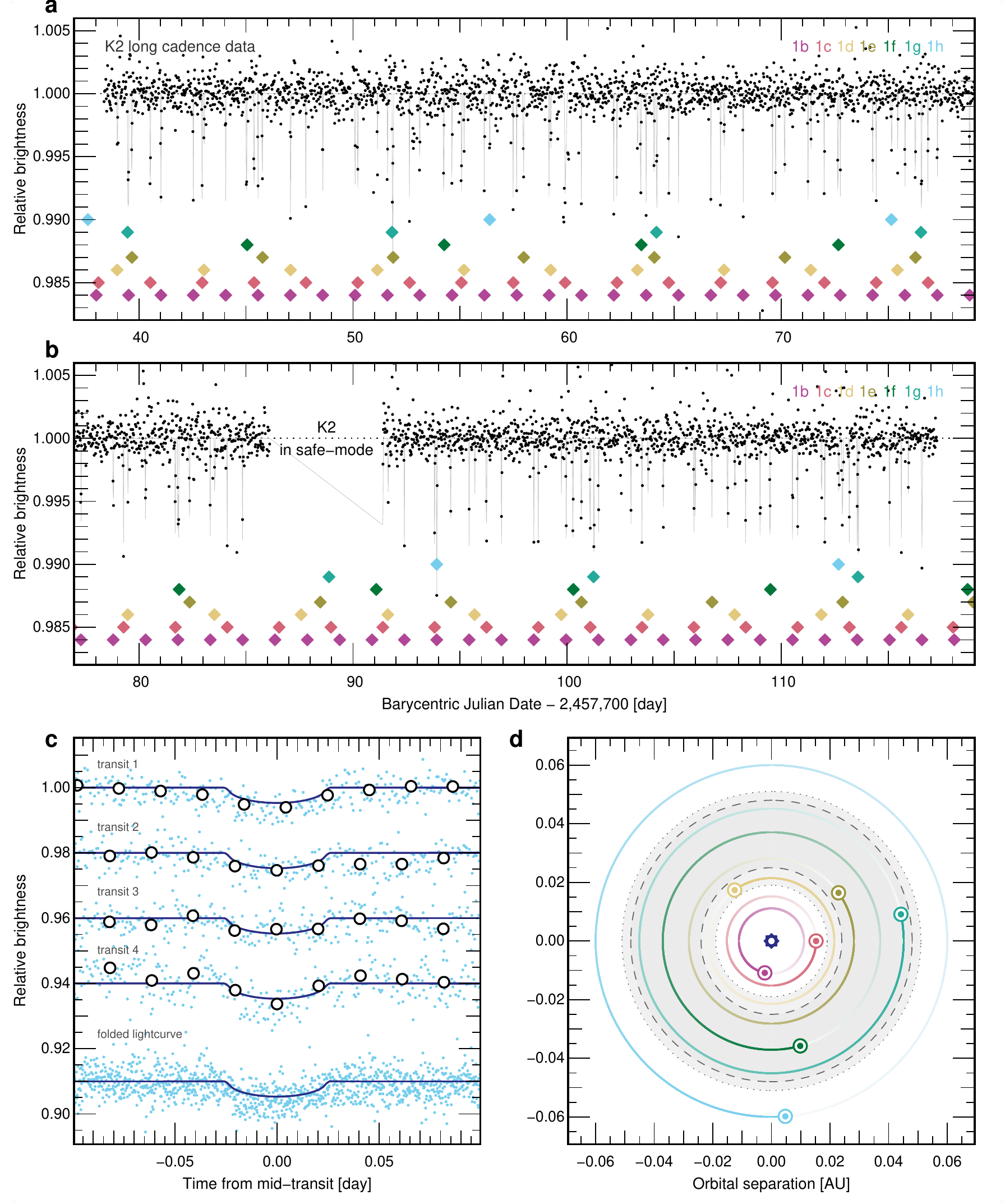}}
    \caption{The long cadence {\it K2} light curve of TRAPPIST-1 detrended with \texttt{EVEREST}. {\bf a, b}: The full detrended {\it K2} light curve with stellar variability removed via LOESS regression (order $=1$; width $=0.15$\,d). Data points are in black, and our highest likelihood transit model for all seven planets is plotted in thin grey. Coloured diamonds indicate which transit belongs to which planet. Four transits of TRAPPIST-1h are observed (light blue diamonds).
    {\bf c} The top four curves show the detrended and whitened short-cadence in light blue, with a transit model based on the {\it Spitzer} parameters in dark blue. Binned data is over-plotted in white for clarity. The folded light curve is displayed at the bottom.
    {\bf d} View from above (observer to the right) of the TRAPPIST-1 system, at the date when the first transit was obtained for this system. The grey region is the surface liquid-water habitable zone.
    \label{fig:fig1-hires}}
\end{figure}

\begin{figure}
    \centerline{\includegraphics[width=0.9\hsize]{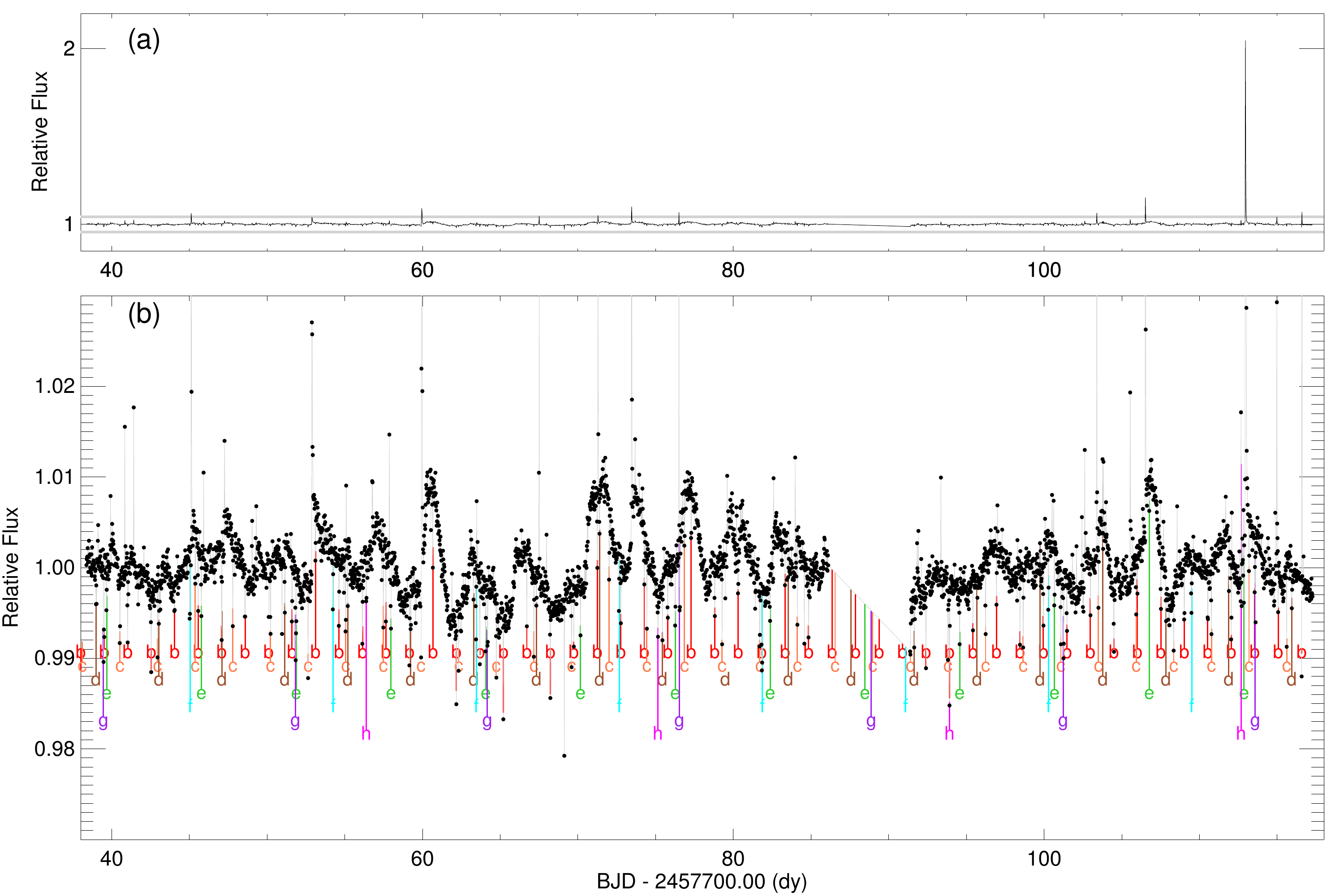}}
    \caption{The entire systematics-corrected {\it K2} dataset with low frequency trends removed. The stellar rotation is apparent in the peaks and troughs of the variability, as are the flares which in some cases appear as single spikes. The planet transits are marked. {\bf a}: Full dynamic range of the curve, including an extreme event at approximately day 113. {\bf b}: Zoomed view of the region outlined in gray in {\bf a}.}
    \label{fig:K2_detrended}
\end{figure}

\begin{table}
\centering
\medskip
\begin{tabular}{|c|c|c|c|c|}
\hline
  planets 1,2,3 & $p$ & $q$ & $\frac{p}{P_1}-\frac{(p+q)}{P_2}+\frac{q}{P_3}$ [d$^{-1}$] &
  $\phi=p\lambda_1-(p+q)\lambda_2+q\lambda_3$\\
\hline
  b,c,d & 2 & 3 & $[-4.6,-0.3]\times10^{-5}$& [$176^\circ$, $178^\circ$]\\
  c,d,e & 1 & 2 & $[-5.2,4.5]\times10^{-5}$& [$47^\circ$, $50^\circ$] \\
  d,e,f & 2 & 3 & $[-1.9,+1.9]\times10^{-4}$&[$-154^\circ$, $-142^\circ$] \\
  e,f,g & 1 & 2 & $[-1.4,+1.1]\times10^{-4}$&[$-79^\circ$, $-72^\circ$] \\
  f,g,h & 1 & 1 & $[-6.0,+0.2]\times10^{-5}$&[$176.5^\circ$, $177.5^\circ$]\\
\hline
\end{tabular}
\caption{The three-body resonances of TRAPPIST-1. The transit times are used to track the $\phi$ angles of each set of three adjacent planets over the dataset, assuming low eccentricities such that transits occur at a phase angle $\lambda=90^\circ$\cite{mills16}. The ranges of three-body frequency and angle given encompass the changes --- most likely librations --- seen during the observations.\label{tab:3body}}
\end{table}

\begin{table}
\centering
\medskip
\begin{tabular}{|l|c|}
\hline
Parameter & Value \\
\hline
Transit depth $(R_p/R_*)^2$ [\%] & 0.346$\pm 0.018$\\[-0.05in]
Transit duration [d] & 0.0525$\pm$0.0008\\[-0.05in]
Impact parameter $b$ [$R_*$] & 0.45$_{-0.08}^{+0.06}$\\[-0.05in]
Mid-transit time, $T_0$ (${\rm BJD_{TDB}}$) & 2,457,662.55284$\pm 0.00037$\\[-0.05in]
Period, P [days] & 18.767$_{-0.003}^{+0.004}$\\[-0.05in]
Radius ratio, $R_p/R_*$ & 0.0588$\pm 0.0016$\\[-0.05in]
Radius, $R_p$ [$R_{\oplus}$] & 0.752$_{-0.031}^{+0.032}$\\[-0.05in]
Inclination, $i$($^\circ$) & 89.76$_{-0.04}^{+0.05}$ \\[-0.05in]
Scale parameter, $a/R_*$ & 109$\pm$4\\[-0.05in]
Equilibrium temperature [K] & 173$\pm$4\\[-0.05in]
Irradiation, $S_p$ [$S_{\rm Earth}$] & 0.165$\pm$0.025 \\[-0.05in]
\hline
Limb-darkening parameters ({\it Kepler} bandpass) & \\
\hline
$u_1$ & 1.00$\pm$0.02\\[-0.05in]
$u_2$ & -0.04$\pm$0.04\\[-0.05in]
\hline
Individual transit timings from K2 (${\rm BJD_{TDB}}$) & \\
\hline
Transit 1 & 2,457,756.3874$_{-0.0013}^{+0.0013}$\\[-0.05in]
Transit 2 & 2,457,775.1539$_{-0.0016}^{+0.0016}$\\[-0.05in]
Transit 3 & 2,457,793.9230$_{-0.0025}^{+0.0024}$\\[-0.05in]
Transit 4 & 2,457,812.6987$_{-0.0042}^{+0.0045}$\\[-0.05in]
\hline
\end{tabular}
\caption{Properties of TRAPPIST-1h, limb-darkening parameters and transit timings derived using a joint {\it Spitzer} and {\it K2} dataset. Parameter values are the medians of the posterior distributions from the MCMC and the associated error bars are the 1-sigma credible intervals. \label{tab:trappist1params}}
\end{table}

%%%
%%%
%%%
%%% METHODS
%%%
%%%
%%%

\clearpage
\begin{methods}
%(3000 words)
\subsection{Light curve preparation.} 
% - Detailed description of the data pre-processing (Rodrigo/Ethan/Marko)
We use the package \texttt{kadenza} to generate a target pixel file (TPF) from the Campaign 12 raw data for TRAPPIST-1 (see Code Availability). In addition to EPIC ID 246199087, which corresponds to a standard-size postage stamp centered on TRAPPIST-1, the {\it Kepler} GO Office also made available a larger, $11\times11$ pixel custom mask with a different ID (200164267), which we use for the purposes of this study. We manually select a rectangular $6\times6$ pixel (24'' $\times$ 24'') aperture centered on TRAPPIST-1 and use the median values of the remaining pixels to perform a column-by-column background subtraction. This process removes dark current and background sky signals while mitigating smear due to bright stars in the same CCD column as the target. Based on simple error function fits to the stellar image, we find that our aperture encloses $> 0.9996$
of the flux of TRAPPIST-1h throughout the entire timeseries.
We then detrend and perform photometry on the pixel-level data using two independent pipelines: \texttt{EVEREST} and a Gaussian process-based detrender. For computational expediency, we perform preliminary analyses on the long cadence data, following up on features of interest in the short cadence data (see below).
% \textbf{Marko's pipeline.}

\subsection{Light curve detrending \--- EVEREST.} 
% - Detailed description of EVEREST (Rodrigo/Ethan)
The \texttt{EVEREST} {\it K2} pipeline\cite{Luger2016} uses a variant of pixel level decorrelation\citeMethods{Deming2015} (PLD) to remove instrumental systematics from stellar light curves. Given a stellar image spread out over a set of pixels $\{\vec{p_i}\}$, \texttt{EVEREST} regresses on polynomial functions of the fractional pixel fluxes, $\vec{p_i} / \sum_j \vec{p_j}$, identifying the linear combination of these that best fits instrumental signals present in the light curve. Because astrophysical signals (such as transits) are equally present in each of the pixels, whereas instrumental signals are spatially variable, PLD excels at removing instrumental noise while preserving astrophysical information. \texttt{EVEREST} uses a Gaussian process (GP) to model correlated astrophysical noise and employs an L2-regularized regression scheme to minimize overfitting.

Since PLD may overfit in the presence of bright contaminant sources in the target aperture\cite{Luger2016}, we manually inspect the TRAPPIST-1 {\it K2} postage stamp and high resolution images of TRAPPIST-1 taken with the APO ARC 3.5m telescope in the SDSS z band to verify that there are no other targets brighter than the background level in our adopted aperture. Given the faint magnitude of TRAPPIST-1 in the {\it Kepler} band, we use the PLD vectors of 14 nearby bright stars (EPIC IDs 246177238, 246165150, 246211745, 246171759, 246127507, 246228828, 206392586, 246121678, 246229336, 246196866, 246217553, 246239441, and 246144695) generated using the same method as above to improve the signal-to-noise ratio of the instrumental model\cite{Luger2017}. We further mask the data in the vicinity of all transits of planets TRAPPIST-1b--g and the potential transits of TRAPPIST-1h when computing the model to prevent transit overfitting. Finally, we divide the light curve into three roughly equal segments and detrend each separately to improve the predictive power of the model. Following these steps, we obtain a detrended light curve for TRAPPIST-1 with a 6-hr photometric precision\citeMethods{Christiansen2012}$^,$\cite{Luger2016} of 281.3\,ppm, a factor of 3 improvement on that of the raw light curve (884.4\,ppm); see Fig.~\ref{fig:fig1-hires}. Before the {\it K2} observation, we estimated the $Kp$ magnitude of TRAPPIST-1 to be $17.2 \pm 0.3$ based on a fit to a corrected blackbody spectrum. However, the photometric precision we achieve with \texttt{EVEREST} is inconsistent with a target dimmer than $Kp \sim 17$. Our detrending therefore suggests that the magnitude of TRAPPIST-1 in the {\it Kepler} band is $16 < Kp < 17$.

\subsection{Light curve detrending \--- Gaussian process (GP) model.} 
% - Detailed description of PIPELINE NAME (Marko)

Independently, we also detrend the data with a GP-based pipeline. To perform aperture photometry, we locate the star using a centroid fit and apply a circular top-hat aperture following the star's centroid coordinates. We then use a GP model to remove the pointing drift systematics
% from the intrinsic variation,
using an additive kernel with separate spatial, time and white noise components\citeMethods{aigrain2015, aigrain2016}:
\begin{equation}
k_{xy}(x_i,y_i,x_j,y_j) = A_{xy}\exp \left[ -\frac{(x_i - x_j)^2}{L_x^2}-\frac{(y_i - y_j)^2}{L_y^2} \right]
\end{equation}
\begin{equation}
k_{xy}(t_i,t_j) = A_{t}\exp \left[ -\frac{(t_i - t_j)^2}{L_t^2} \right]
\end{equation}
\begin{equation}
 K_{ij} = k_{xy}(x_i, y_i, x_j, y_j) + k_t(t_i, t_j) + \sigma^2 \delta_{ij}
\end{equation}
where $x$ and $y$ are the pixel coordinates of the centroid, $t$ is the time of the observation, and the other variables ($A_{xy}$, $L_x$, $L_y$, $A_t$, $L_t$, $\sigma$) are hyperparameters of the GP model.
We use the \texttt{GEORGE} package\citeMethods{ambikasaran2014} in \texttt{PYTHON} to implement the GP model. To find the maximum likelihood hyperparameters we use a differential evolution algorithm\citeMethods{Storn1997}, followed by a local optimization. This method was tested on magnitude $16-18$ stars observed in Campaign 10 of {\it K2}, and we use the results of those tests, and of previous GP applications to {\it K2} data\citeMethods{aigrain2016}, to inform our priors on the hyperparameters. 

For the TRAPPIST-1 data, we use an iterative sigma-clipping method to remove outliers and prevent the time component from overfitting. This method has been previously used in the \texttt{k2sc}\citeMethods{aigrain2016} pipeline. First, using fiducial hyperparameter values based on analysis of a Campaign 10 target, we remove all measurements with residuals greater than $3\sigma$ from the mean GP prediction. With the remaining measurements, we update the hyperparameters by maximizing the GP likelihood. Using these parameters, we once again clip all $3\sigma$ outliers and maximize the GP likelihood using only the remaining measurements. The final detrending is calculated for all points, including outliers. 

\subsection{Photometric analysis I.} 
% - Rodrigo's search in short cadence
We fold the long cadence data on the dynamically predicted orbital period of 18.765\,d for TRAPPIST-1h with a time of first transit constrained by the {\it Spitzer} observation, revealing a feature consistent with a transit in both the \texttt{EVEREST} and GP-based light curves. In order to confirm the planetary nature of this signal, we analyze the short cadence data. To this end, we use \texttt{kadenza} to generate a short cadence TPF of TRAPPIST-1 and detrend it in windows of 1.5--2\,d centered on each of the four features using \texttt{EVEREST}. We use the PLD vectors of five bright stars observed in short cadence mode (EPIC IDs 245919787, 246011640, 246329409, 246331757, and 246375295) to aid in the detrending. When generating these light curves, we explicitly mask large flares so that these do not inform the fit. Following this procedure, we obtain binned 6-hr photometric precision of 266.6, 176.1, 243.4, and 219.3\,ppm in each of the four windows. The short and long cadence data in these windows is shown in Supplementary Fig.~1. For transits 3 and 4, additional correction of the light curve is necessary, since the transit of TRAPPIST-1h coincides with a transit of TRAPPIST-1b (panel 3a) and a small flare (panel 4b). The transit of TRAPPIST-1b is subtracted out using a transit model\citeMethods{MandelAgol2002} with the {\it Spitzer} parameters and a mid-transit time determined from the data, yielding the light curve in panel 3b. The flare is fit using a 3-parameter flare model\citeMethods{Davenport2014} for stars observed with {\it Kepler}, yielding the light curve in panel 4b (see also Supplementary Fig.~2). In both cases, the transit of TRAPPIST-1h is visible in the residuals. In Supplementary Fig.~3 we show the folded short cadence data after accounting for TTVs, with a transit model based solely on the {\it Spitzer} parameters.

\subsection{Photometric analysis II.} 
% - Detailed description of the analysis codes (Laetitia, Brice, Michael)
We use the long-cadence detrended light curve to perform a transit search with a box least-squares fitting algorithm (BLS)\citeMethods{Kovacs:2002}. We set the BLS to orbital periods ranging from 10\,d to 50\,d. The ratio of the transit duration over the planet orbital period is set between 0.0007 and 0.06 to include a wide range of orbital periods, eccentricities, and impact parameters for additional planets in the system. Due to \emph{Kepler's} 30-minute cadence, transits of planets orbiting TRAPPIST-1 appear significantly smeared out, which we take into account in our analysis. The highest peak in the periodogram corresponds to a signal with a 15.44-day period. Its origin stems from residuals in blended transits with TRAPPIST-1c and d and from one outlier in the data. No signal is seen at the two other epochs where a transit should have appeared, which confirms the signal being spurious. The next highest peak in the periodogram corresponds to a $\sim$18.77-day period and a transit center at
%
%HJD=7756.39 
% RL changed this for consistency with the figures and text
%
$\mathrm{BJD} = 2,457,756.39$
that is consistent with the single transit seen with {\it Spitzer}. 

We then use a Markov Chain Monte Carlo (MCMC) algorithm previously described in the literature\citeMethods{Gillon:2012} to derive the transit parameters of TRAPPIST-1h from the detrended light curve. Each photometric data point is attached to a conservative error bar that accounts for the uncertainties in the detrending process presented in the previous section. We impose normal priors in the MCMC fit on the orbital period, transit mid-time center and impact parameter for planets TRAPPIST-1b to g to the values recently published\cite{Gillon2017}. We further assume circular orbits for all planets\cite{Gillon2016,Gillon2017}. We also include normal priors for the stellar properties, which are $\mathcal{N}$(0.080, 0.007$^2$)\,M$_{\odot}$ for the mass, $\mathcal{N}$(0.117, 0.004$^2$)\,R$_{\odot}$ for the radius, $\mathcal{N}$(2,555, 85$^2$)\,K for the effective temperature and $\mathcal{N}$(0.04, 0.08$^2$)\,dex for the metallicity\cite{Gillon2017}. We use these stellar parameters to compute the quadratic limb-darkening coefficients $u_1$ and $u_2$ in the {\it Kepler} bandpass from theoretical tables\citeMethods{Claret:2011}. 
In a first MCMC fit, we use a 7-planet model that includes all seven planets, with no prior information on the orbital period or $t_0$ of TRAPPIST-1h. In the second fit we employ a 6-planet model that excludes TRAPPIST-1h. We use the results from both MCMC fits to compute the Bayesian and Akaike Information Criteria (BIC and AIC, respectively) to determine which model is favoured. We find BIC values of 2888 and 2897 for the 7 and 6-planet models, respectively. This corresponds to a Bayes Factor $e^{(BIC_1-BIC_2)/2}=90$ in favour of the 7-planet model. Similarly, we find AIC values of 2691 and 2725 for the 7 and 6-planet models, respectively. We perform a third MCMC fit to refine the transit parameters of TRAPPIST-1h. For this fit, we use as input data the {\it K2} short-cadence data centered on the 4 transits of TRAPPIST-1h (Fig.~\ref{fig:fig1-hires}) and the single transit light-curve previously obtained with {\it Spitzer}. This fit includes a model for TRAPPIST-1b and a flare that both affect the transit shape of TRAPPIST-1h in the K2 short-cadence data. This fit also allows for TTVs for the individual transit timings. We find photometric precisions of 365 ppm and $\sim$1100 ppm per 10 min for {\it Spitzer} and {\it K2} respectively. We report the median and 1-sigma credible intervals of the posterior distribution functions for the transit parameters of TRAPPIST-1h in Table~\ref{tab:trappist1params}, along with the individual transit times.

\subsection{Photometric analysis III.} 
% - Rodrigo's joint instrumental/transit model description
In order to prevent the overfitting of transit features, we mask all transits of $b-h$ when detrending with \texttt{EVEREST}. However, this inevitably results in a lower detrending power during transits. A powerful alternative to the detrend-then-search method employed above is to simultaneously fit the instrumental and transit signals, without masking those features\citeMethods{Foreman-Mackey2015}. We therefore conduct a second, separate blind search on the \texttt{EVEREST} light curve specifically for TRAPPIST-1h. Given a raw light curve $\vec{y}$, a data covariance matrix $\mathbf{\Sigma}$, and a single-transit model $\vec{m}_{t_0}$ centered at $t = t_0$, the log likelihood of the transit fit is

\begin{equation}
    \log\mathcal{L} = -\frac{1}{2}(\vec{y}-\vec{m}_{t_0})^\top\mathbf{\Sigma^{-1}}(\vec{y}-\vec{m}_{t_0}) + C
\end{equation}

\noindent where $C$ is a constant. The data covariance matrix, $\mathbf{\Sigma}$, is the sum of the astrophysical covariance and the L2-regularized PLD covariance and is given by

\begin{equation}
    \mathbf{\Sigma} = \mathbf{X}\mathbf{\Lambda}\mathbf{X}^\top + \mathbf{K}
\end{equation}

\noindent where $\mathbf{K}$ is the astrophysical covariance given by the \texttt{EVEREST} GP model, $\mathbf{X}$ is the matrix of PLD regressors (the design matrix), and $\mathbf{\Lambda}$ is the prior covariance of the PLD weights (the regularization matrix) which we obtain by cross-validation\cite{Luger2017}. Since the transit shape and duration of TRAPPIST-1h are known\cite{Gillon2017}, the only free parameter in the search was $t_0$, the time of transit. We therefore evaluate $\vec{m}_{t_0}$ multiple times, centering the transit model at each long cadence and computing the likelihood of the transit model fit as a function of cadence number. We then subtract these values from the log likelihood of the data with no transit model ($\vec{m}_{t_0} = 0$) and multiply by 2 to get the delta-chi squared ($\Delta\chi^2$) metric, which measures the decrease in the $\chi^2$ value of the fitted light curve for a transit of TRAPPIST-1h centered at each cadence. Finally, we also compute $\Delta\chi^2$ conditioned on the known ``true'' transit depth of TRAPPIST-1h, $d_0 = 0.00352 \pm 0.000326$:

\begin{equation}
    \Delta\chi^2_\mathrm{cond} = \Delta\chi^2 - \left(\frac{d - d_0}{\sigma_d}\right)^2
\end{equation}

\noindent where

\begin{equation}
    d = \sigma_d^2 \vec{m}_{t_0}^\top\mathbf{\Sigma^{-1}}\vec{y}
\end{equation}

\noindent is the maximum likelihood depth of the transit model and

\begin{equation}
    \sigma_d^2 = (\vec{m}_{t_0}^\top\mathbf{\Sigma^{-1}}\vec{m}_{t_0})^{-1}
\end{equation}

\noindent is the variance of the depth estimate. Positive peaks in $\Delta\chi^2$ indicate features that are well described by the transit model, while positive peaks in $\Delta\chi^2_\mathrm{cond}$ reveal features that are well described by the transit model with depth $d = d_0$. In Supplementary Fig.~4 we show the two $\Delta\chi^2$ metrics across the full TRAPPIST-1 light curve after subtracting a transit model for planets $b-g$ based on their {\it Spitzer} parameters. The strongest features in the $\Delta\chi^2$ plot (top) are flares, as these can be fitted out with an inverted transit model. When conditioning on the true depth of TRAPPIST-1h (bottom), the significance of most of the flare features decreases, revealing the four peaks of TRAPPIST-1h (red arrows).

In order to assess the robustness of our detection, we compute the total $\Delta\chi^2_\mathrm{cond}$ as a function of orbital period. Starting from the time of transit of TRAPPIST-1h in the {\it Spitzer} dataset, we compute the times of transit in the {\it K2} light curve for 500,000 values of the orbital period evenly spaced between 1 and 50\,d and sum the values of $\Delta\chi^2_\mathrm{cond}$ at each transit time to produce the total $\Delta\chi^2_\mathrm{P}$. We sum these in two different ways. First, we linearly interpolate the grid of $\Delta\chi^2_\mathrm{cond}$ to each transit time to obtain $\Delta\chi^2_\mathrm{P}$ for perfectly periodic transits. Next, to allow for TTVs of up to one hour, we take the largest value of $\Delta\chi^2_\mathrm{cond}$ in the vicinity of each transit time and sum them for each period. We adopted a tolerance of 2 cadences, corresponding to maximum TTVs of 1\,hr. Our results are shown in Supplementary Fig.~5, where we plot the periodic $\Delta\chi^2_\mathrm{P}$ (top) and the $\Delta\chi^2_\mathrm{P}$ allowing for TTVs (bottom). In both cases, the period of TRAPPIST-1h (18.766\,d) and its $1/2$ and $1/3$ period aliases emerge as the three strongest peaks. The peak at 18.766\,d is the strongest signal in the period range constrained by the {\it Spitzer} transit\cite{Gillon2017} and confirms our detection of TRAPPIST-1h.

%\subsection{Dynamical simulations.}
% - TTVs, Architecture and long-term stability of the system. (Eric, Sean, Brice, Dan, Dan)
%We revisit the masses of the planets using transit-timing based on the new transit times found with the {\it K2} dataset. Our prior analysis neglected planet $h$, but with its newly measured transit times, we now add it to the timing analysis. Given the large number of planets in the system, we continue to assume plane-parallel orbits for the planets, which yields five free parameters per planet, $(t_{0,i},P,m_i/m_*,e_i\cos{\omega_i},e_i\sin{\omega_i})$, for a total of 35 free parameters describing the system. To further constrain the orbits of the planets and break degeneracies between the eccentricity vectors and masses\cite{Lithwick2012,Deck2015}, we place a Gaussian prior upon each component of the eccentricity vector of
% each planet, allowing the width of the Gaussian, $\sigma_e$, to vary as a free parameter in the analysis; we assume that a single value of $\sigma_e$ applies to all seven planets.  % To describe:  TTV model (refs), linearization, mixture model (?), results of TTV analysis, TTV plots & model
 
\subsection{Three-body angles.} 
The mean longitude of a planet with orbital period $P$ is an angular variable that progresses at a constant rate with respect to time $t$, which is measured from the time the planet passes a given reference direction:
\begin{equation}
\lambda = \frac{360^\circ}{P} t,
\end{equation}
with $\lambda$ measured in degrees.
For transiting planets, the reference direction is taken as the plane perpendicular to the observer's line of sight, as the planet is progressing towards the transiting configuration. We assume orbits with negligible eccentricities\cite{mills16}, for which $\lambda = 90^\circ$ at transit mid-time, so that we may write
\begin{equation}
\lambda=360^\circ \left(\frac{1}{4} + \frac{t-T_n}{P}\right)
\end{equation}
for each planet, where $T_n$ is the time of transit of the $n^\mathrm{th}$ planet. For a 3-body resonance $(p,q)$, we may therefore express the three-body angle as
\begin{equation}
\phi = 360^\circ \left[-p \frac{T_1}{P_1} + (p+q) \frac{T_2}{P_2} - q \frac{T_3}{P_3} + p \frac{t}{P_1} - (p + q)\frac{t}{P_2} + q\frac{t}{P_3} \right].
\end{equation}
The state of a $\phi$ value is assessed when individual transit times of three planets are taken near each other. For instance, for the transit times\cite{Gillon2017} $T_f=7662.18747$, $T_g=7665.35151$, $T_h=7662.55463$, and $(p,q)=(1,1)$, we compute $\phi=177.4^\circ$ at $t=7664$ ($\mathrm{BJD}-2,450,000$). 

\subsection{Tidal simulations.}
Tidal interactions with the star are important for all 7 planets. We perform N-body simulations of the system including an equilibrium tidal dissipation formalism\citeMethods{eggleton98,leconte10} using the  Mercury-T code\cite{Bolmont2015}. 
We use orbital parameters from the discovery paper\cite{Gillon2017} and a period of TRAPPIST-1h of 18.765\,d (a near 2:3 resonance configuration with planet $g$). 
%This configuration is compatible with our new estimate.
We consider the planets' spins to be tidally synchronized with small obliquities. 
This is justified because even if the age of the system is 400\,Myr (a lower estimate for the age of TRAPPIST-1),
%\textbf{[CITATION NEEDED]} -- It's ok, do it in review
planetary tides would have had time to synchronize the spins\citeMethods{leconte2015} (even when atmospheric tides are accounted for).
We test different initial eccentricities and different values for the planets dissipation factors (from 0.01 to 10 times the Earth's value\citeMethods{DeSurgyLaskar1997}).

Our simulations show that the planets' orbital eccentricities are likely to be low. In just a few Myr all eccentricities decrease to below 0.01 for dissipation factors $\geq0.1$ times the Earth's value.
Due to planet-planet interactions, the eccentricities do not decrease to zero but instead reach an equilibrium value determined by the competition between tidal damping and planet-planet eccentricity excitation\citeMethods{Bolmont2013}.
All planets stay in resonance during the evolution towards tidal equilibrium.
The small equilibrium eccentricities are sufficient to generate significant tidal heating. 
Assuming the TRAPPIST-1 planets have a tidal dissipation equal to the Earth's, we find that TRAPPIST-1b might have tidal heat flux similar to Io's\citeMethods{Spencer2000} ($\sim3$\,W\,m$^{-2}$), with peaks at more than 10\,W\,m$^{-2}$ (corresponding to $\sim10^4$\,TW) when the eccentricity oscillation is at its maximum. 
Planets $c$ through $e$ have tidal heat fluxes higher than Earth's internal (primarily radioactive) heat flux\citeMethods{Davies2010} ($\sim$0.08\,W\,m$^{-2}$) but lower than Io's. 
TRAPPIST-1f, g and h have a tidal heat flux inferior to Earth's.
Supplementary Fig.~6 shows a possible snapshot of the system's evolution over the course of 40\,yr.
%SML: Is that supposed to say 40 years...?? 
This very high flux could plausibly generate intense volcanism on the surfaces of the inner planets, with potential consequences for their internal structures.

\subsection{Planet habitability.}
We calculate the minimum stellar flux required for liquid water with the LMD 1D/3D Global Climate Model\citeMethods{Turbet2016} using a synthetic spectrum of TRAPPIST-1 based on its reported $T_\mathrm{eff}$, $\log g$, metallicity, and bolometric luminosity\cite{Gillon2017}, obtaining a value of 300 $\mathrm{W/m^2}$, which is 100 $\mathrm{W/m^2}$ higher than the planet's present-day instellation. Our results are in agreement with habitable zone boundaries computed for a 3000\,K star\citeMethods{Kopparapu2013}. Assuming zero albedo, we find that the equilibrium temperature of TRAPPIST-1h is 169$\pm 4$ K.

Whether or not TRAPPIST-1h presently hosts an atmosphere is unclear. Given its radius measurement and a range of possible compositions (from pure water ice to pure iron), the mass of TRAPPIST-1h is likely in the range 0.067--0.863~$M_\oplus$; if TRAPPIST-1h has an Earth-like composition, this value is 0.33~$M_\oplus$\citeMethods{Sotin2007}. Assuming TRAPPIST-1h migrated to its current location quickly, the planet's low surface gravity could have led to vigorous hydrodynamic escape of a primordial atmosphere in the first few 100 Myr after its formation, since at that time TRAPPIST-1 was significantly brighter and TRAPPIST-1h would have been interior to the habitable zone\citeMethods{Luger2015a,Luger2015b}. The presence of a subsequently outgassed atmosphere, however, cannot be ruled out.

In theory, the surface of TRAPPIST-1h could harbor liquid water under such an outgassed atmosphere if it is H$_2$-rich. Atmospheres made of H$_2$\citeMethods{Stevenson1999,Pierrehumbert_Gaidos2011}, N$_2$-H$_2$\citeMethods{Wordsworth_Pierrehumbert2013} or CO$_2$-H$_2$\citeMethods{Ramirez_Kaltenegger2017,Wordsworth_2016} have been shown to provide a sufficient greenhouse effect and internal heat blanketing for even lower instellation levels. Unless prevented by high altitude clouds, transit spectroscopy with the \emph{Hubble} Space Telescope may be able to reveal or rule out such H$_2$-rich atmospheres.

\subsection{Stellar variability.}
To establish the rotation period of TRAPPIST-1, we clean the long cadence \texttt{EVEREST} light curve (Fig.\ \ref{fig:K2_detrended}) of remaining outliers, transits, and flares. We iteratively fit and remove low frequencies that remain in the light curve from the detrending process. To extract the rotation period, we calculate the Discrete Fourier Transform (DFT) following a previously established method,\citeMethods{Kurtz85} and proceed to fit the detected frequency with non-linear least-squares. The rotation period is determined to be $3.30\pm0.14$\,d. 

To determine the occurrence rate of stellar flares, we again take the \texttt{EVEREST} detrended long cadence light curve and remove transits and outliers. We require flares to have peak emission 1\% above the normalized continuum flux and two consecutive signals above the continuum. We detect a total of 19 flare events, corresponding to an average rate of 0.26 d$^{-1}$. A more refined determination of flaring rate and energies requires analysis of the short cadence data, which will be presented in a forthcoming paper.   

%%%
%%%
%%%
%%% CODE AND DATA AVAILABILITY
%%%
%%%
%%%

\subsection{Code availability.} The \texttt{kadenza} code we use to generate pseudo-target pixel files for all {\it K2} targets was downloaded from \url{https://doi.org/10.5281/zenodo.344973}.
The code we use to generate and analyze the \texttt{EVEREST} light curves for TRAPPIST-1 is openly available at \url{https://github.com/rodluger/trappist1}. A static version of the repository has been archived at \url{https://doi.org/10.5281/zenodo.437548}.

\subsection{Data availability.} The $K2$ raw cadence data used in this study is available for download at \url{https://archive.stsci.edu/missions/k2/c12_raw_cadence_data/}. The pseudo-target pixel files for TRAPPIST-1 and its neighboring stars generated with \texttt{kadenza} are archived at \url{https://doi.org/10.5281/zenodo.437876}. The detrended TRAPPIST-1 long cadence light curve and segments of the short cadence light curve in the vicinity of the transits of TRAPPIST-1h are available at \url{https://doi.org/10.5281/zenodo.437548}.
All other data that support the plots within this paper and other findings of this study are available from the corresponding author upon reasonable request.

\end{methods}

%%%
%%%
%%%
%%% REFERENCES FOR METHODS
%%%
%%%
%%%

\bibliographystyleMethods{naturemag}
\bibliographyMethods{k2_trappist1_methods}

%%%
%%%
%%%
%%% SUPPLEMENTARY FIGURES AND TABLES
%%%
%%%
%%%

\supplement

\thispagestyle{empty}
\begin{figure}
\centerline{\includegraphics[width=0.9\hsize]{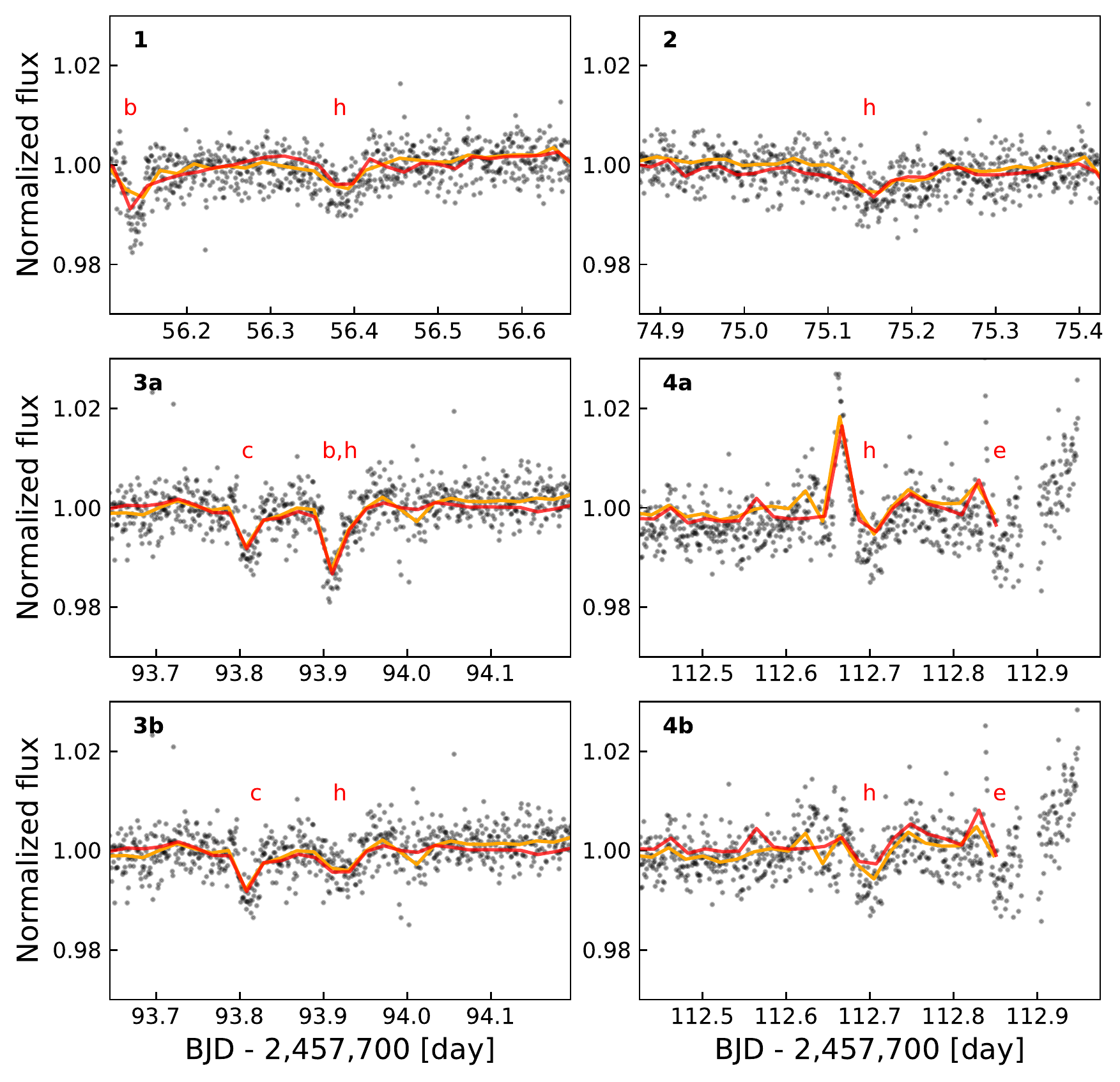}}
\caption{The four transits of TRAPPIST-1h. The detrended short cadence data is shown as black dots. Orange curves are these data binned to 30-minute cadence, and red curves are the detrended long cadence data. The transits of TRAPPIST-1h and other planets are indicated with red letters. Corrections have to be made to remove the simultaneous transit of $b$ in transit 3 and a near-simultaneous flare in transit 4. The uncorrected data for these transits is shown in the middle row, and the data with these features removed is shown in the bottom row.
\label{fig:trappist1h}}
\end{figure}

\thispagestyle{empty}
\begin{figure}
\centerline{\includegraphics[clip,trim=0cm 0cm 0cm 1cm,width=0.9\hsize,]{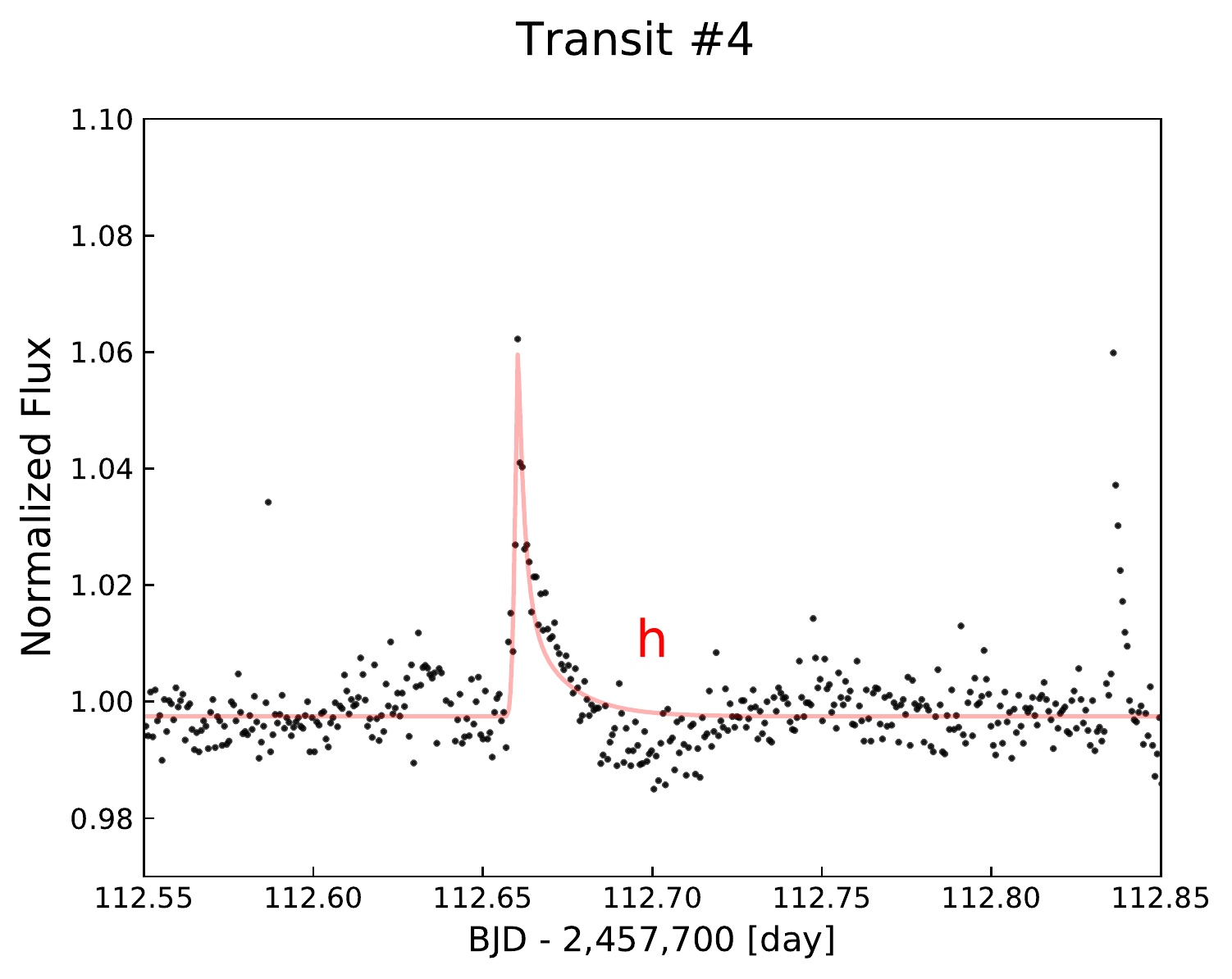}}
\caption{The short cadence data in the vicinity of transit 4 of planet TRAPPIST-1h, where a small flare is visible. The transit of TRAPPIST-1h occurs ${\sim}60$ short cadences (1 hr) after the peak of the flare. A least-squares fit to the flare is shown in red; the data during the transit of TRAPPIST-1h is clearly lower than the baseline.
\label{fig:flarefit}}
\end{figure}

\thispagestyle{empty}
\begin{figure}
\centerline{\includegraphics[width=\hsize]{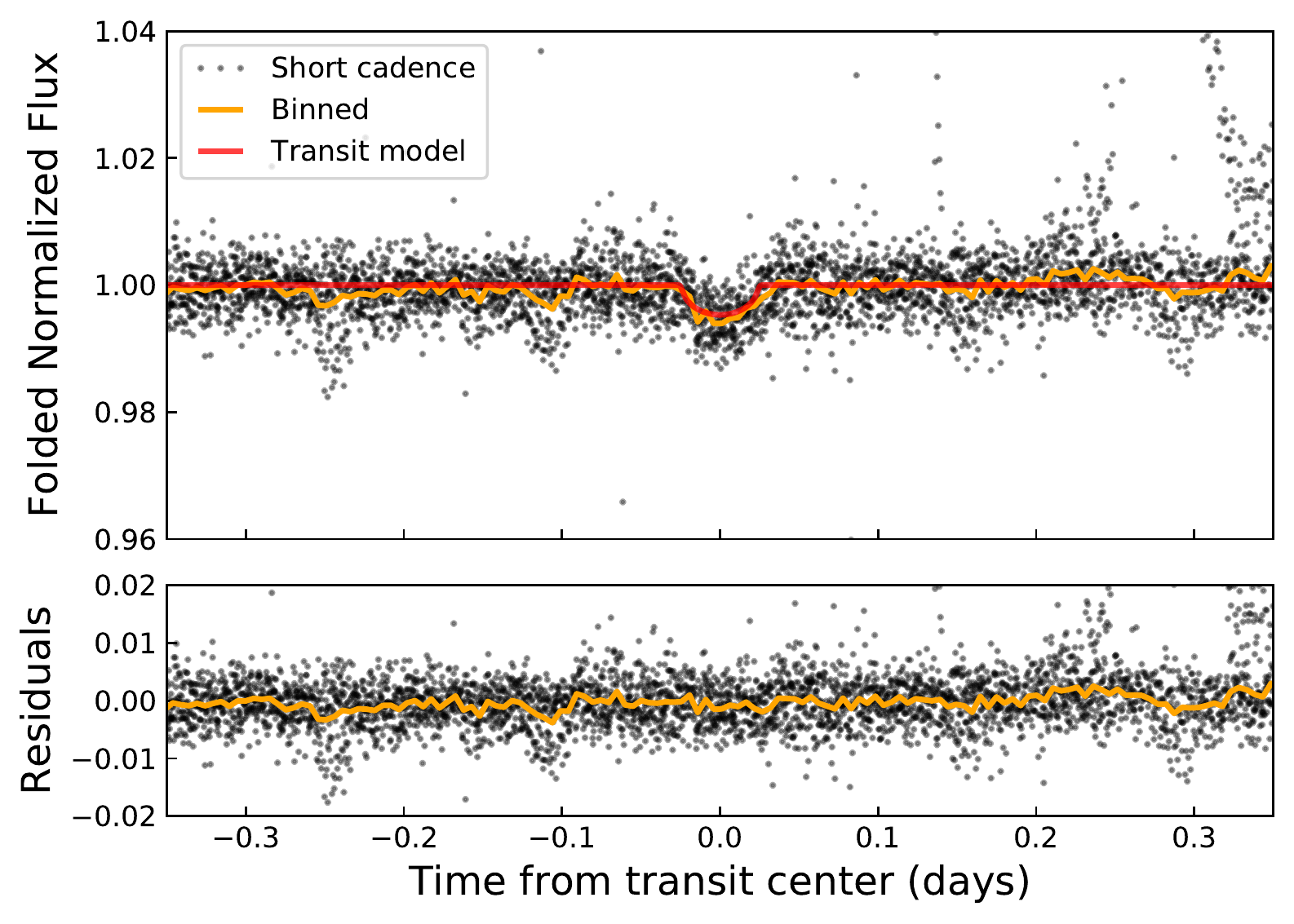}}
\caption{The short cadence data folded on the four transits of TRAPPIST-1h after correcting for TTVs and subtracting a simultaneous transit of TRAPPIST-1b and a near-simultaneous flare. Other transits of TRAPPIST-1b to g have not been removed and are visible in parts of the data. The data downbinned by a factor of 30 is shown as the orange line, and a transit model based {\it solely} on the {\it Spitzer} parameters is shown in red. The residuals (data minus this model) are shown at the bottom.
\label{fig:trappist1hfold}}
\end{figure}

\thispagestyle{empty}
\begin{figure}
\centerline{\includegraphics[width=\hsize]{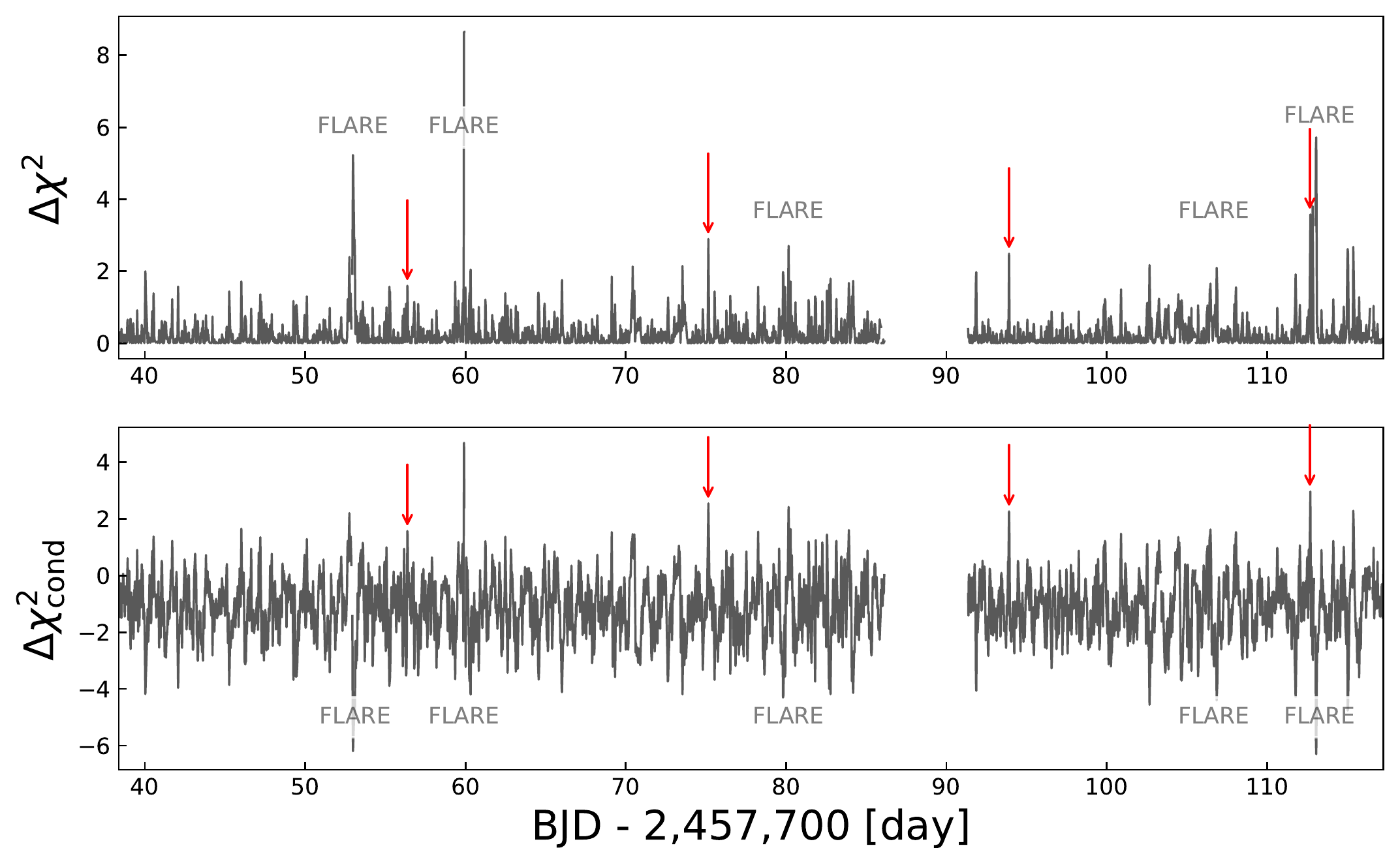}}
\caption{Delta-chi squared (top) and delta chi-squared conditioned on the true depth (bottom) for each long cadence in the TRAPPIST-1 \texttt{EVEREST} light curve. The four transits of TRAPPIST-1h are indicated with red arrows. In the top plot, spikes appear at the location of several flares, as these can be fit with inverted transits. Conditioning on the true depth (bottom) removes many of these features by penalizing transit depths that are inconsistent with the observed {\it Spitzer} value.
\label{fig:everest_deltachisq}}
\end{figure}

\thispagestyle{empty}
\begin{figure}
\centerline{\includegraphics[width=\hsize]{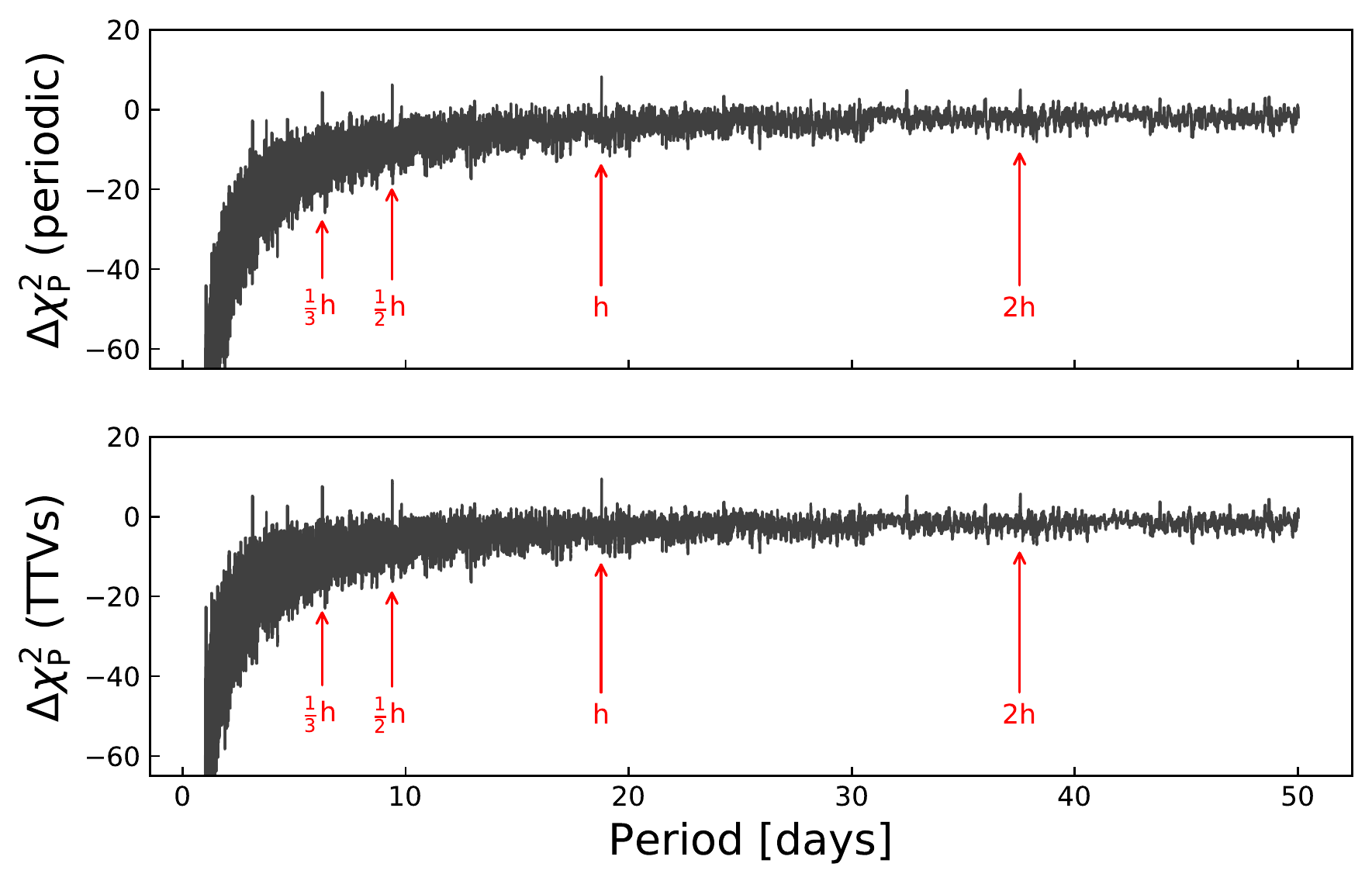}}
\caption{Delta-chi squared values as a function of orbital period for TRAPPIST-1h, taking the {\it Spitzer} transit time and assuming perfectly periodic orbits (top) and allowing for TTVs of up to 1\, hr (bottom). The period of TRAPPIST-1h (18.766\,d) and its aliases emerge as the strongest peaks in both plots.
\label{fig:everest_powerspec}}
\end{figure}

\thispagestyle{empty}
\begin{figure}
\centerline{\includegraphics[width=0.9\hsize]{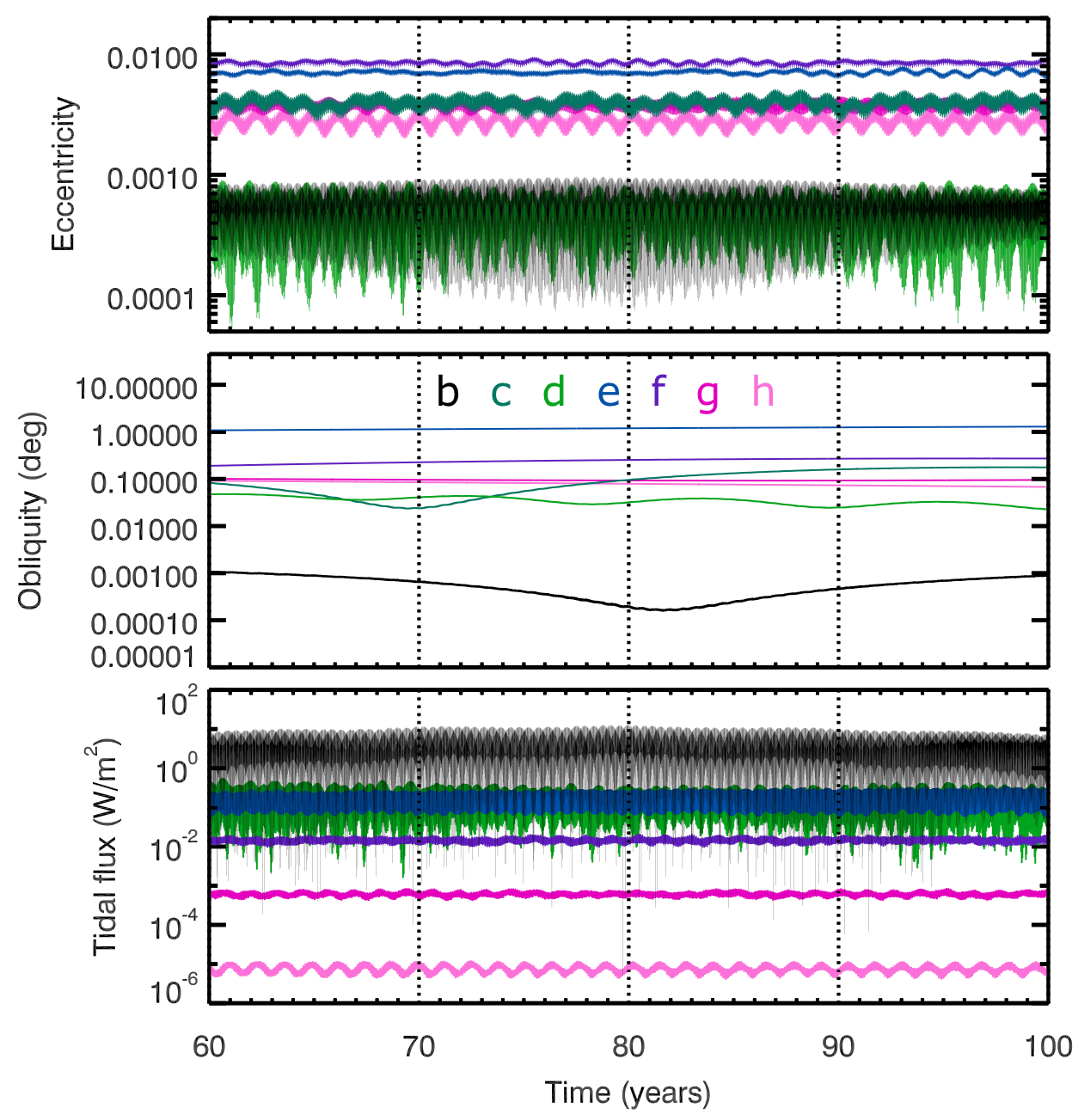}}
\caption{Possible short term evolution of the eccentricity, obliquity and tidal heat flux of the TRAPPIST-1 planets. The different planets are represented by different colors from black (planet b) to light pink (planet h).\label{tides}}
\end{figure}

\thispagestyle{empty}

\end{document}